\documentclass{PoS}

\def\br{\begin{eqnarray}}
\def\er{\end{eqnarray}}
\def\be{\begin{equation}}
\def\ee{\end{equation}}

\title{QCD phenomenology with infrared finite SDE solutions}

\ShortTitle{QCD phenomenology with SDE solutions}

\author{\speaker{A. A. Natale} \\ 
        Instituto de F\'{\i}sica Te\'orica - UNESP\\
        Rua Dr. Bento T. Ferraz,271 - Bloco II, S\~ao Paulo, SP, Brazil\\ 
        E-mail: \email{natale@ift.unesp.br}}

\abstract{Recent progress in the solution of Schwinger-Dyson equations (SDE), as well as lattice simulation
of pure glue QCD, indicate that the gluon propagator and coupling constant are infrared (IR)
finite. We discuss how this non-perturbative information can be introduced into the QCD perturbative
expansion in a consistent scheme, showing some examples of tree level hadronic reactions that successfully fit
the experimental data with the gluon propagator and coupling constant depending on a dynamically generated gluon mass. This infrared mass scale acts as a natural cutoff and eliminates some of the {\it {ad hoc}} parameters usually found 
in perturbative QCD calculations. The application of these IR finite Green's functions in the case of higher order terms
of the perturbative expansion is commented.}

\FullConference{International Workshop on QCD Green's Functions, Confinement and Phenomenology\\
		 September 7-11, 2009\\
		 ECT$^*$, Trento, Italy}

\begin{document}

\section{Introduction}

Many years ago Mandelstam obtained a solution of the SDE for the gluon propagator, in the Landau gauge and in the case of pure gauge QCD, that behaved as $1/k^4$ in the infrared \cite{mand}. This solution was named as a ``confined gluon solution" because it naturally leads to a linear confining potential. This result motivated intense
phenomenological studies where the IR enhanced propagator was substituted by more tractable functions, however these
functions were still peaked at origin of momenta. Examples of this program can be found in the review of Ref.\cite{rob}.

Although not widely diffused, it is mentioned in Mandelstam's paper about the possible existence of a massive solution for the gluon propagator, but this solution was discarded since the beginning in his approach. Within the same approximation \cite{an1} but with one improved three gluon vertex function \cite{an2} it can be shown that this solution indeed exists in the Landau gauge. Of course, such soft IR behavior certainly requires a more subtle confinement explanation \cite{cornwall}. 

The problem with a large part of the more recent SDE calculations for the gluon propagator 
(for a review, see \cite{r2}) and the ones
discussed above is that they do not lead to transverse solutions as required by gauge invariance, and it seems that the only way to have the Schwinger's mechanism ({\it gauge invariant}) realization in non-Abelian gauge theories is rearranging the SDE diagrams through the pinch technique \cite{cornwall,bp}. This was, for the first time, devised by Cornwall in 1982 \cite{cornwall}, who 
obtained a gauge invariant solution for the gluon propagator that behaves as $1/[k^2 + m^2(k^2)]$. As $k^2 \rightarrow 0$ the function $m^2(k^2)$ was interpreted as a dynamical gluon mass with the limit $m^2(k^2\rightarrow 0) = m^2_g$.
In this picture the coupling constant is also IR finite with a fixed point behavior described by
\be
{\bar{\alpha}}_{sd} (0) \equiv \frac{1}{4\pi b \ln [(4m_g^2)/\Lambda^2]} \,\, ,
\label{eq1}
\ee
where $\Lambda = \Lambda_{QCD}$ is the QCD scale where the perturbative coupling becomes singular.
Recent lattice QCD simulations present clear evidence for the 
dynamical generation of a gluon mass (a long list of references about these simulations can be found
in Ref.\cite{cornwall2} and a recent simulation can be seen in Ref.\cite{bogo}), 
with the SDE result for a massive gluon fitting nicely the lattice data \cite{abp}.

There are several reasons to review the strong interaction phenomenological calculations
on the light of IR finite gluon propagator and coupling constant. A simple one is that in this
scheme we get rid of the Landau singularity in the coupling constant, which introduce singularities in the physical amplitudes that do not correspond to the expected physical behavior. There are also singularities in
QCD amplitudes, like the one in the two-gluon QCD Pomeron model \cite{lan}, that disappear when dealing with a
dynamically generated gluon mass. The dynamical gluon mass also introduces a natural
IR cutoff which may substitute the one that is always present in many perturbative QCD calculations. Moreover it is usually argued that
the perturbative QCD series can be reorganized in order to ameliorate its behavior, but, most important of all,
this optimization of the perturbative expansion is quite dependent on the infrared
behavior of the coupling constant \cite{dok}, and the dynamical mass generation scheme is precisely giving
us one hint of which direction to go in order to improve the perturbative series. Finally, besides all these
reasons, the fact that IR finite Green's functions with a dynamically generated gluon mass scale provide a better agreement between experiment and theory, as will be shown here,
corroborates the SDE and lattice results.

As we shall discuss in the next section, when presenting the expressions for the gluon propagator and
coupling constant, the SDE solutions cannot determine the $m_g$ value, and the best we can do is to
determine the ratio $m_g/\Lambda$. We will show that several perturbative calculations can be improved
with the knowledge of propagator and coupling in the full range of momenta, and in some cases the experimental
data can only be fitted with the help of these IR finite quantities. In Section 2 we discuss that the relevant
scheme to introduce these non-perturbative information into the QCD perturbative expansion is the one
named Dynamical Perturbation Theory (DPT), then we present examples of phenomenological calculations
that make use of IR finite Green's functions. Different observables are computed as a function
of $m_g$ and all data indicate a small range for the dynamical gluon mass, what is impressive if we
consider that our examples involve different hadronic mass scales or wave functions. In Section 3 we comment how this procedure can be extended to compute higher order terms of the perturbative expansion and we make a summary in Section 4.

\section{DPT at tree level} 

\subsection{DPT and IR finite SDE solutions }

A prescription of how the non-perturbative SDE solutions can be inserted into the
perturbative QCD expansion was proposed by Pagels and Stokar many years ago, in the approach denominated DPT \cite{pagels}. In their
scheme the amplitudes that do not vanish to all orders in perturbation theory are given by their free
field values, while amplitudes that vanish as $\lambda\propto e^{-1/g^2}$ are retained, and possibly
dealt with in an expansion in $g^n\lambda$. The work of Ref.\cite{pagels} was particularly concerned with the effect of
a dynamically generated quark mass, but as we now know from the SDE solutions that the gluon and coupling constant also have an infrared finite value, we can extend their formulation and generalize the perturbative expansion
using the quark and gluon propagators with a dynamical mass and the IR finite coupling constant. This means
that we should perform perturbation theory with the dressed quark and gluon propagators and the effective 
charge (dependent on the gluon mass).

The SDE solutions and the lattice results were discussed at length in this workshop \cite{aptr}, therefore
we will not enter into details about the solutions and will just present the gluon propagator and coupling constant in
the case that QCD generates a dynamical gluon mass, otherwise we mention in the references
where a different SDE solution is used.
We consider a gluon propagator that will have the form
\be
\imath {{\Delta}}_{\mu\nu} (q)=P_{\mu\nu} {\Delta}(q)+\xi \frac{q_\mu q_\nu}{q^4} \,\, ; \,\,  P_{\mu\nu}=-g_{\mu\nu} + \frac{q_\mu q_\nu}{q^2} \,\, ,
\ee
where ${\Delta}(q)$ is the gauge invariant scalar part of the gluon propagator, which in Euclidean space has the form
\be
{\Delta}(Q^2)\propto \frac{1}{Q^2 + m_g^2 (Q^2)} \,\, .
\ee
The gluonic SDE solutions allow us to write a new propagator ${\hat{\Delta}}^{-1}(Q^2)$ which absorbs all the renormalization group logs,
exactly as happens in QED with the photon self-energy, and form the product ${\hat{d}}(Q^2)=g^2{\hat{\Delta}}(Q^2)$ which is
a renormalization group invariant.
The dynamical gluon mass ($m_g^2(Q^2)$) is given by a complicated expression falling with the momentum as $1/Q^2$ \cite{ap}, and 
it can be simply approximated by \cite{an2}
\be
m_g^2(Q^2)=\frac{ m_g^4}{Q^2 + m_g^2 } \,\, ,
\ee
where $m_g \approx {\cal{O}}(1.2-2)\Lambda$, with $\Lambda = \Lambda_{QCD}\approx 300 \,$ MeV. As this is still a complicated
expression to take into account when calculating loops, in practical calculations, the best we can do is to assume $m_g^2(Q^2)\approx m_g^2$. 
A simple fit for the coupling constant that is factored out in this procedure is given by \cite{cornwall}
\be
{\bar{\alpha}}_{sd} (q^2) = \frac{1}{4\pi b \ln [(4m_g^2 -q^2 -\imath\epsilon )/\Lambda^2]} \,\, ,
\label{eq31}
\ee
where $b = (33-2n_f)/48\pi^2$.  
Eq.(\ref{eq31}) clearly shows the existence of the IR fixed-point shown in Eq.(\ref{eq1}).
It must be stressed that the fixed point shown in Eq.(\ref{eq1}) does not depend on a specific process, 
it is uniquely obtained as we fix $\Lambda$ and, 
in principle, it should be exactly determined if we knew how to solve QCD. The quantity ${\hat{d}}(Q^2)$ is the one that appear in
all loop calculations. As in QED, where the only ultraviolet divergences are associated with the vacuum polarization, and affect
the renormalization of the coupling constant, in QCD, as the pinch technique is applied and the vacuum polarization is summed, the only difference is that instead of a renormalized coupling and a massless propagator it is a factor like ${\hat{d}}(Q^2)$ that appears in the calculation,
indicating the existence of a massive propagator and an IR finite charge. 

There are several examples of the use of DPT with the propagator and coupling constant discussed above \cite{several}
apart the ones that will be commented in the next subsections \cite{pion,tcross,bmeson,pomer}, where the nice description of the experimental
data within this procedure can be perceived.  

\subsection{Pion form factor}

\begin{figure}[htbp]
\vspace{-1.0cm}
\begin{center}
\includegraphics[width=9cm]{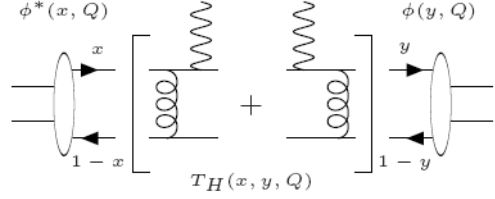}
\vspace{-1.5cm}
\caption{The leading-order diagrams that contribute to the pion form factor.}
\label{figfpiqcd}
\end{center}
\end{figure}

\begin{figure}[htbp]
\begin{center}
\includegraphics[width=8cm]{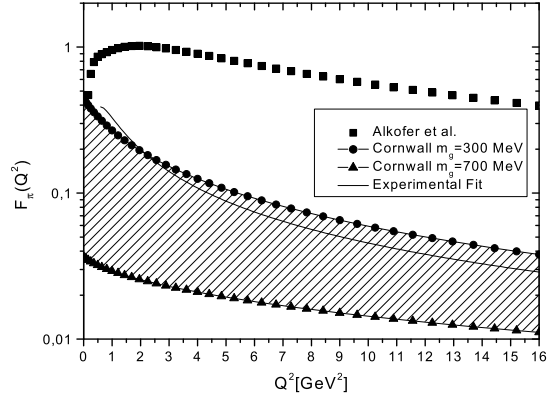}
\caption{Comparison of the experimental data for $F_{\pi}$(solid line) with the theoretical
determination of the form factor via DPT for a range of dynamical gluon masses. We also show the results obtained with the SDE solution of Ref.\cite{r2}.}
\label{fpiex}
\end{center}
\end{figure}

The asymptotic pion form factor is predicted in perturbative QCD,
according to Brodsky and Lepage \cite{bl}, as
\be 
F_{\pi}(Q^2)= \int_{0}^{1}\!\!dx \! \int_{0}^{1}\!\!dy \,
\phi^{*}
(y,\tilde{Q}_y) T_H(x,y,Q^2) \phi(x,\tilde{Q}_x) \,\, , 
\label{fpiy}
\ee
where $\tilde{Q}_x =min(x,1-x)Q$ and $Q$ is the 4-momentum in Euclidean
space transferred by the photon. The function $\phi(x,\tilde{Q}_x)$ 
is the pion wave function, that gives the amplitude
for finding the quark or antiquark within the pion carrying
the fractional momentum $x$ or $1-x$, respectively. The function $T_H(x,y,Q^2)$, is the hard-scattering
amplitude that is obtained by computing the quark-photon
scattering diagrams shown in Fig.(\ref{figfpiqcd}). 
The perturbative kernel $T_{H}$ is giving by
\be
 T_{H}(x,y,Q) = \frac{64\pi}{3Q^2}\left\{\frac{2}{3}
\frac{\alpha_s[(1-x)(1-y)Q^2]}{(1-x)(1-y)} 
+ \frac{1}{3} \frac{\alpha_s(xyQ^2)}{xy} \right\} \,\, ,
\label{k1}
\ee
while in the case of DPT the gluon propagator and coupling constant are exchanged by the dressed
(non-perturbative) functions:
\be
T_H(x,y,Q^2) = \frac{64\pi}{3}\left[
\frac{2}{3}{\bar{\alpha}}_{sd}(K^2)\Delta(K^2) +
\frac{1}{3}{\bar{\alpha}}_{sd}(P^2)\Delta(P^2)\right] \,\, ,
\ee
where $K^2=(1-x)(1-y)Q^2$ and $P^2=xyQ^2$.
The details of the calculation, as the dependence on the pion wave functions and other 
quantities, can be found in Ref.\cite{pion}. The comparison of the theoretical result with
the experimental data can be seen in Fig.(\ref{fpiex}). The important point to notice is
that this result comes out from a convolution of the gluon propagator and coupling constant
with the pion wave functions, and, as happens in this case, for each observable that we shall comment in this work there will be
different wave functions or scales involved in the calculation, however all the experimental
data will indicate the same range of values for the dynamical gluon mass, which points to
a strong phenomenological constraint on the SDE solution that we are using. 
Notice that the purely perturbative result, obtained with the kernel giving by Eq.(\ref{k1}), does not show such a nice fit for the experimental data.

\subsection{Hadronic cross section in a QCD-inspired model}

According to a QCD-inspired model to compute hadronic scattering \cite{blo}, the cross section for producing jets with $p_{T}>p_{T_{min}}$ (through the dominant process $gg\to gg$) is proportional to
\be
\sigma_{jet}(s) = \int_{p^2_{T_{_{min}}}} dp^2_T \,
 \frac{d\hat{\sigma}_{gg}}{dp^2_T} 
\int_{x_1 x_2 > 4p^2_T /s} dx_1 dx_2 \, g(x_{1},Q^2) \, g(x_{2},Q^2)
\label{jetx}
\ee
where $g(x,Q^2)$ is the gluon flux, and a minimum transversal momentum ($p^2_{T_{min}}$) establish
the region where perturbation theory can be applied. The standard calculation assumes the following 
elementary partonic cross section for gluon-gluon scattering
\be
\hat{\sigma}_{gg}(\hat{s})  \propto \frac{9\pi \alpha_0^2}{m_0^2} \, \theta (\hat{s} -m_0^2) \,\, , 
\label{q321}
\ee
where $m_{0}$ and $\alpha_0$ are fitted parameters.

\begin{figure}[htbp]
\begin{center}
\includegraphics[width=8cm]{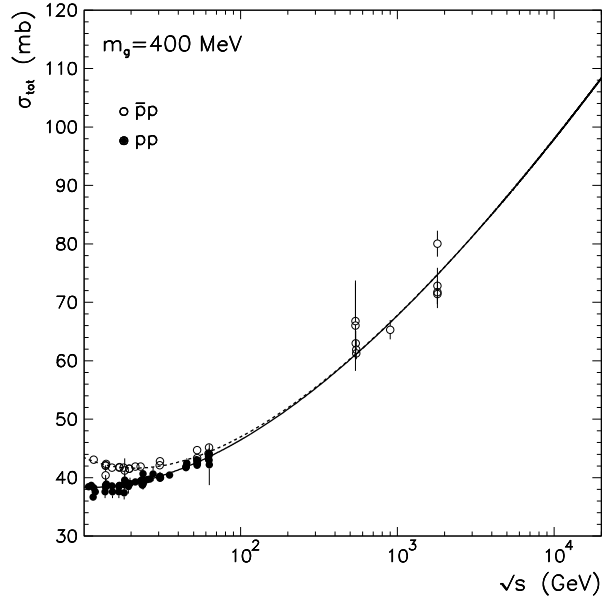}
\caption{Total cross section for $pp$ (solid curve) and $\bar{p}p$ (dashed curve) scattering.} \label{difdad2}
\end{center}
\end{figure}

DPT allow us to compute elementary cross section, like $\hat{\sigma}_{gg}(\hat{s})$ with IR finite quantities,
where $m_{0}$, $\alpha_0$ and $p^2_{T_{min}}$ are substituted by $m_g$!
We fitted the $pp$ and $p\bar{p}$ scattering data keeping $m_g$ as a free parameter.
Taking a $5\%$ variation on the minimal $\chi^{2}/DOF$ value indicate $m_{g}\approx 400^{+350}_{-100}$ MeV
(for details, see Ref.\cite{tcross}). One of the fits to the experimental data can be seen in Fig.(\ref{difdad2}).
Again, even considering the quite different calculations of Eq.(\ref{fpiy}) and Eq.(\ref{jetx}), it is amazing
how the observables that we discussed up to now lead to the same gluon mass range, supporting
the DPT scheme based on IR finite SDE solutions depending on a dynamical gluon mass. Moreover, the two different fitting parameters of Eq.(\ref{q321}) were eliminated from the calculation in our procedure!

\subsection{Non-leptonic annihilation B meson decays}

\begin{figure}[htbp]
\begin{center}
\includegraphics[width=8cm]{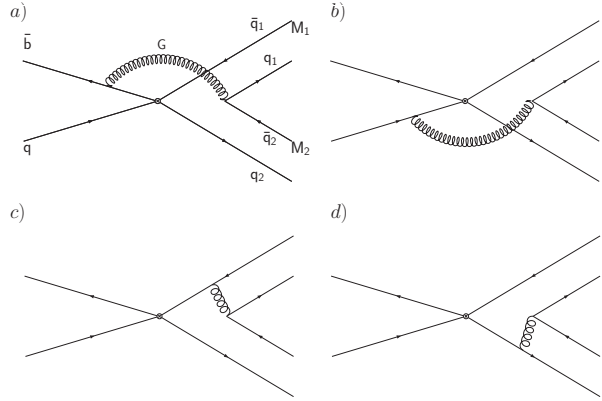}
\caption[dummy0]{Perturbative diagrams leading to two-body non-leptonic annihilation B decays}
\label{fig1}
\end{center}
\end{figure}

The use of DPT can be exemplified by the diagrams of Fig.(\ref{fig1}), which show the different contributions
for the two-body non-leptonic annihilation B meson decays in the factorization approach. DPT predicts that to each gluon exchange depicted in the different
pictures will correspond to a dressed gluon propagator that enters into the amplitude calculation, as 
well as its coupling constant, i.e. the perturbative $\alpha_s$ and gluon propagator were substituted by
the product ${\hat{d}}(Q^2)=g^2{\hat{\Delta}}(Q^2)$. In the standard perturbative QCD calculation we
are faced with end-point divergences due to soft gluon emission and an arbitrary cutoff is introduced
in the amplitude:
\be
\int \frac{dx}{x} = \ln\frac{m_B}{\Lambda_h}(1+\rho e^{i\phi})  \,\,\, , \,\,\,\, 0\leq \rho\leq 1 \,\, .
\label{eqc}
\ee 
This divergence is eliminated in the DPT scheme with IR finite Green's functions.

Some branching ratios of these two-body non-leptonic annihilation B meson decays are shown in Table (\ref{table1}).
The results of Table (\ref{table1}) contain only the leading twist contribution, and we may expect an increase in the branching ratio
values when the higher twist contributions are added. A discussion on the meson wave functions dependence as well as on the choice of B meson
scales ($\mu=m_b$ or $\mu=m_b/2$) can be found in Ref.\cite{bmeson}. The important point is that the best description of the data
is giving by an IR gluon propagator and coupling constant that can be associated to a dynamically generated gluon mass. A full picture of these
decays calculated in a totally perturbative approach is not so compelling as the one presented here, besides the fact that, although small, there
is a dependence on the {\it {ad hoc}} cutoff discussed in Eq.(\ref{eqc}).
\begin{table*}[htb]
\caption{Branching  ratios for non-leptonic annihilation $B$  decays obtained
  with the infrared  finite gluon propagator and coupling constant discussed in the beginning of Section 2. These values
  were obtained with $m_g=500$MeV. A 
  complete list of the results and experimental data can be found in Ref.\cite{bmeson}.  }
\label{table1}
\newcommand{\m}{\hphantom{$-$}}
\newcommand{\cc}[1]{\multicolumn{1}{c}{#1}} 
\renewcommand{\tabcolsep}{2pc} 
\renewcommand{\arraystretch}{1.2} 
\begin{tabular}{@{}lll}
\hline 
 Decay channels & $m_g = 500$MeV & $Experiment$ \\ 
 \hline
${\mathcal Br}(B_s^0 \rightarrow \pi^+\pi^-)\times 10^7$ & 1.58& $< 13.6$ \\
${\mathcal Br}(B_d^{0} \rightarrow K^+K^-)\times 10^8$ &  7.18 &  $4\pm 15\pm 8$\\
${\mathcal Br}(B_s^0 \rightarrow D^-\pi^+)\times 10^6$ & 1.54 & -- \\
${\mathcal Br}(B_s^0 \rightarrow D^+\pi^-)\times 10^7$ &  1.88& -- \\
${\mathcal Br}(B_d^0 \rightarrow D_s^-K^+)\times 10^5$ &  1.98& $2.9\pm 0.4\pm 0.2$ \\
${\mathcal Br}(B^-_d \rightarrow D_s^+K^-)\times 10^8$ & 0.99 & $< 1.1 \times 10^5$\\
\hline
\end{tabular}\\[2pt]
\end{table*}

\subsection{A QCD-Pomeron model}

Hadronic cross sections can also be computed in the scope of Regge theory, where the amplitude for elastic proton-proton scattering at high energy is dominated by the Pomeron. Landshoff and Nachtmann (LN) proposed a
QCD-Pomeron model where the basic Pomeron structure is represented by two non-perturbative gluon exchange, where the gluon has a finite correlation length \cite{lan}. 
\begin{figure}[htbp]
\begin{center}
\includegraphics[width=8cm]{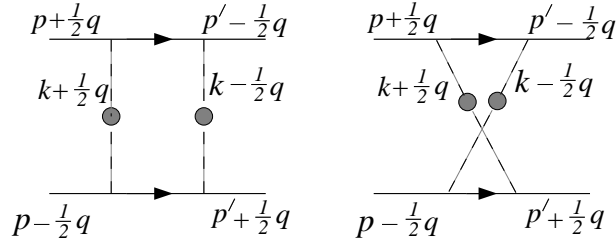}
\caption[dummy0]{$T_{1}$ and $T_{2}$. Two-gluons exchange model for the Pomeron.}
\label{t1t2}
\end{center}
\end{figure}
The differential proton-proton cross section can be writing as
\be
  \frac{d\sigma}{dt}= \frac{|A(s,t)|^2}{16\pi s^2} \,\, , \nonumber
\ee
where the amplitude $A(s,t)$ is giving by the diagrams shown in
Fig.(\ref{t1t2}). 
\vspace{1.0cm}
\be
A(s,t)= \imath s 8 \alpha_{s}^2 \left[ T_{1} - T_{2} \right] \,\, . \nonumber
\ee

\begin{figure}[htbp]
\begin{center}
\includegraphics[width=8cm]{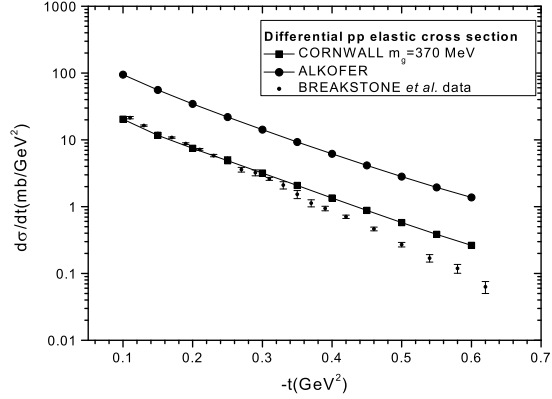}
\caption{Differential $pp$ elastic cross section at
$\sqrt(s)=53\, \mbox{GeV}$ computed within the
Landshoff--Nachtmann model for the Pomeron,
using different infrared couplings and gluon propagators
obtained from DSE solutions.} \label{transition}
\end{center}
\end{figure}

Several results obtained for the LN model in the DPT scheme can be found in Ref.\cite{pomer} as well as in some of the references \cite{several}. Actually the LN proposal match exactly with the DPT ideas, demanding the introduction of an IR finite non-perturbative gluon propagator and coupling constant.
We just show the result for the differential proton-proton elastic cross section at $\sqrt(s)=53\, \mbox{GeV}$
depicted in Fig.(\ref{transition}). 
In Fig.(\ref{transition}) we compare different SDE solutions and the data is better explained by
a gluon propagator and coupling constant associated to a gluon mass scale of ${\cal{O}}(400)\, \mbox{MeV}$.
Notice that the two gluon exchange model can explain the elastic scattering only at small $t$. For 
large transferred momentum it is necessary to add a three gluon exchange with opposite parity. Again
we do have different wave functions and scales that have to be introduced in the calculation (see Ref.\cite{pomer}), but
we still have a good agreement with the data for the same range of dynamical gluon masses. 

\section{DPT at higher orders of the perturbative expansion}
 
The question that we would like to discuss in this section is that if 
we can have precise phenomenological tests of the $\alpha_s$ and gluon propagator infrared behavior
at higher orders of the perturbative expansion. This type of test together with the ones discussed
in the previous section can definitively provide a strong test for the SDE solutions.
These SDE solutions can be applied in the DPT scheme at the loop level, however this is not a
trivial matter and requires the use of the pinch technique \cite{bp} to disentangle the different 
contributions that come from different Green's functions inside loops. What we will present
here is a preliminary result of the DPT application in the case of the Bjorken sum rule and many 
aspects of this calculation are still in progress \cite{anp}.
  
It has been pointed out that the Bjorken 
sum rule \cite{bj} could be used to know the $\alpha_s$ behavior up to low energies. This is, for instance, 
the point of view followed in Ref.\cite{deur2} using the idea of an effective coupling, and also of 
Ref.\cite{milton} in the case of analytic perturbation theory. Therefore this will be a perfect
arena to test the behavior of the infrared quantities that we discussed up to now. 
The polarized Bjorken sum rule can be written as \cite{bj}
\be
\Gamma^{p-n}_1 (Q^2) = S_{Bj}= \int^{1}_{0} \, dx \,  \left[ g_1^p (x,Q^2)- g_1^n (x,Q^2) \right] \, ,
\label{eq21}
\ee
where $g_1^{p}(g_1^n )$ is the first spin structure function for the proton (neutron), which
were measured recently at quite low $Q^2$ \cite{deur}.

The QCD correction to this sum rule up to the fourth order in the strong coupling constant $\alpha_s$ for massless particles and 
effective number of flavors $n_f=3$ is given by \cite{verm}
\br
\Gamma^{p-n}_1 (Q^2) = S_{Bj} (Q^2) &=& \frac{1}{6} \left|\frac{g_A}{g_V} \right| \left[1-\frac{\alpha_s(Q^2)}{\pi} -
3.58\left(\frac{\alpha_s(Q^2)}{\pi}\right)^2 \right. \nonumber \\
&& \left. - 20.21 \left(\frac{\alpha_s(Q^2)}{\pi}\right)^3 - 
130.0 \left(\frac{\alpha_s(Q^2)}{\pi}\right)^4 + ... \right] \,\, , 
\label{eq22}
\er
where $g_A$ and $g_V$ are constants appearing in the nucleon beta decay. In principle the experimental data on the structure functions can be used to determine $\alpha_s (Q^2)$.

\begin{figure}[h]
\centering
\includegraphics[width=0.6\columnwidth]{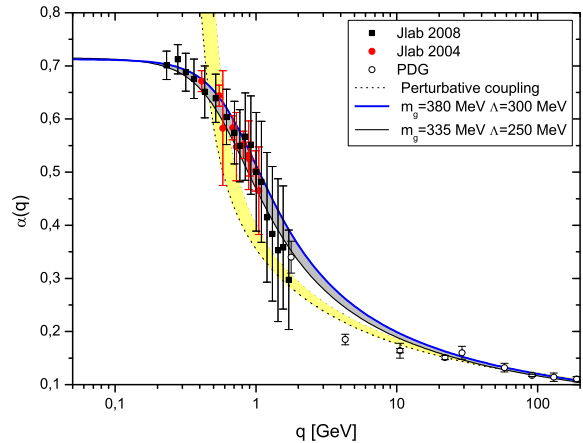}
\caption[dummy0]{ Effective coupling constant extracted from the experimental data of $\Gamma^{p-n}_1 (Q^2)$.}
\label{fig4}
\end{figure}

In our preliminary calculation we have not considered the effect of dynamical masses inside the loops (we expect 
that at 
leading order this effect is small). We also neglect power corrections to the Bjorken sum rule, because many calculations
have shown that their effect is negligible, but we also believe that the power corrections will be softened
by the infrared finite gluon propagator behavior \cite{anp}. 
Eq.(\ref{eq22}) was solved up to order ${{\alpha}}^4$ in order to obtain the value of the gluon mass scale using the experimental data of $\Gamma^{p-n}_1 (Q^2)$ obtained at Jlab \cite{deur}. We show our result in Fig.(\ref{fig4}), which is fitted by Eq.(\ref{eq31}) with a dynamical gluon mass value equal to $m_g = {\cal{O}}(300-400)\,$MeV, that is compatible with previous phenomenological determinations of $m_g$ \cite{several}. 
The match between the effective charge ${\bar{\alpha}}_{sd}$ with the experimental data is impressive.
We stress that any other different behavior for the infrared coupling constant visibly do not show such a
nice fit for the experimental data. There are many points that still need analysis in this procedure: a) It would be
interesting to have a formal demonstration of the DPT scheme realization in the case of the Bjorken
sum rule through the use of the pinch technique, b) Dynamical gluon and quark masses, even if their effect
are small inside loops, should be considered in future work. They should improve the match in Fig.(\ref{fig4}) of the 
experimental data with the non-perturbative coupling in the $1$ GeV region and c) Higher order corrections should
be fully calculated in the DPT scenario in order to confirm our order of magnitude estimates. These calculations
certainly will be quite difficult but they are necessary once we consider the good description of the experimental
data shown in Fig.(\ref{fig4}).

\section{Summary}

SDE solutions as well as lattice simulation
of pure glue QCD are indicating that the gluon propagator and coupling constant are infrared
finite. In Ref.\cite{abp} we can see a nice agreement between these non-perturbative methods, where
the lattice data for the gluon propagator is fitted by a SDE solution associated with a dynamically
generated gluon mass. We show that several strong interaction observables computed within perturbative
QCD improved by the knowledge of the Green's functions, at the full range of momenta in the DPT scheme, also
provide strong support for the SDE and lattice results. The full scenario works so well
that we can barely neglect the possible existence of the dynamical mass generation mechanism in QCD.

All the examples of hadronic phenomenology that we have discussed are well described by a gluon
propagator and strong coupling constant that are dependent on a dynamically generated gluon mass.
No matter we deal with improved perturbative QCD calculations or QCD inspired models we verify that the
experimental data is fitted with a dynamical gluon mass scale $m_g \approx ${\cal{O}}$(2\Lambda_{QCD})$.
In many cases this mass scale helped us to reduce the number of arbitrary parameters in the
calculations. The tests that we performed are non-trivial in the sense that they result from the
calculation of physical quantities where the gluon propagator or product of propagators are integrated
weighted by different functions (involving different mass scales), and all quantities
show agreement with the experimental data for gluon masses that are in the same
range of masses predicted by Cornwall several years ago \cite{cornwall}. It is hard to believe that
such coincidence is a fortuitous one. 

A preliminary account of a phenomenological test of SDE solutions at the loop level was discussed in Section 3.
The simple analysis introduced there is giving a signal that the DPT procedure, with the inclusion of IR
finite Green's functions, is quite promising, and details about this approach shall be presented elsewhere \cite{anp}.             

\section*{Acknowledgments}
I would like to thank the organizers of the ``International Workshop on QCD Green's Functions, Confinement and Phenomenology" for the invitation and quite pleasant organization. The Conselho Nacional de Desenvolvimento Cient\'{\i}fico e Tecnol\'ogico (CNPq-Brazil) and Funda\c c\~ao de Amparo \`a Pesquisa do Estado de S\~ao
Paulo are gratefully acknowledged for financial support.

\end{document}